\documentstyle[prb,floats,aps,twocolumn,psfig]{revtex}

\begin{document}

\draft 
\tightenlines
\widetext

\title{Correlation calculations for the reconstruction of
the Si (100) surface}

\author{Beate Paulus} 
\address{Max-Planck-Institut f\"ur
Physik komplexer Systeme, N\"othnitzer Stra\ss e 38, D-01187 Dresden, Germany}
\maketitle

\begin{abstract}
Ab initio multi-reference configuration interaction calculations are performed for the 
Si(100) surface using a cluster approach. The convergence with
respect to the cluster size is checked and the final results are taken from 
a ${\rm Si}_{32}{\rm H}_{28}$ cluster which models two dimers and
six bulk layers. We find for the ideal as well as for the 
p($1\times 2$) reconstruction a singlet ground state 
consisting of several configurations. The energy gain due to 
forming the symmetric dimer in the p($1\times 2$) structure is 1.75 eV, 
the bond length of the dimer is 2.35 \AA\  which is very close to the
bulk value. In contradiction to the LDA results and in agreement with previous
correlation calculations we do not find an asymmetric  p($1\times 2$) structure. 
\end{abstract} 
\vspace*{2cm}

\narrowtext

\section{Introduction}
\begin{figure}
\psfig{figure=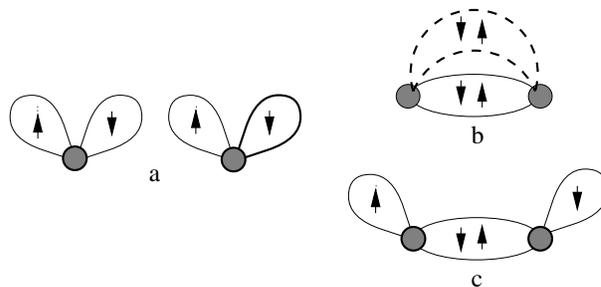,width=8cm,angle=-90}
\caption{\label{structure}The figures show sketches
of the orbital structure of the ideal surface (a),
and of a closed-shell (c) and open shell (b) structure
of the reconstructed surface.} 
\end{figure}

The Si(100) surface is one of the best examined, both 
experimentally
\cite{aono82,yang83,holland84,jedrecy90,badt94,bullock95,felici97} 
and theoretically\cite{verwoerd80,redondo82,artacho90,roberts90,weakliem91,dabrowski92,jing92,tang92,fritsch95,krueger95,ramstad95,yang97}, 
semiconductor surfaces
due to its great technological importance for semiconductor
devices.
LEED experiments\cite{yang83,holland84}, X-ray diffraction 
measurements\cite{jedrecy90,felici97},
scanning tunneling microscopy (STM)\cite{badt94}
and surface core-level photoelectron diffraction experiments\cite{bullock95}
find down to temperatures of 90K that on 
the Si(100) surface alternating buckled dimers in
p(2$\times$2) or c(4$\times$2) structure are formed. These experimental findings 
are in agreement with those of various density functional calculations (DFT) applying
the local density approximation (LDA) to a slab model of the 
surface\cite{roberts90,dabrowski92,fritsch95,ramstad95}.
Applying the LDA to cluster models of this surface\cite{tang92,yang97}
the energy difference between the buckled reconstruction and a
reconstruction forming symmetric dimers is very small, so they
can not draw a concluding result on the ground state structure.
Earlier work done with semi-empirical quantum chemical 
methods\cite{verwoerd80} favours the asymmetric reconstruction as
a closed-shell Hartree-Fock treatment\cite{redondo82} does.
Including electronic correlations at the
generalized valence bond level\cite{redondo82} or within a
configuration interaction (CI) approach\cite{jing92} yields a symmetrically
dimerized ground state similar to the findings of unrestricted Hartree-Fock
calculations\cite{artacho90}.\\
The wide spread density functional methods rely on the 
ground-state density and do not provide a many-body wave function.
The electronic correlations are included implicitly only via the 
exchange-correlation 
functional. Regarding the process forming a dimer on the surface
the question arises whether DFT can cope with bond breaking and bond formation.
In the ideal Si(100) surface each surface atom has
two singly occupied $sp^3$ hybrids (see Fig.\ref{structure} a).
In a simple-minded picture the surface energy is lowered by
reducing the number of singly occupied orbitals. A bond is formed
between two surface atoms yielding the (1$\times$2) dimer structure.
The nature of the bond can have two limiting cases: If a double-bond
is generated between the two Si atoms (see Fig.\ref{structure} b),
this would lead to a closed-shell structure of the surface. Or,
a single-bond is generated leaving  on each surface atom a
singly occupied orbital (see Fig.\ref{structure} c).
The true electronic structure of the surface can only be solved
within a multi-determinant approach, which can describe all 
configurations between the two limiting cases.
A second advantage
of this approach is that the ideal and reconstructed surface
are described at the same level of accuracy because we apply no
restrictions on the wave function as single-reference approaches do.  
Up to now the multi-reference methods were applied to adsorption and
diffusion processes of hydrogen on Si(100)\cite{wu92}.
The most sophisticated correlation method applied to  
the reconstruction itself is by Jing and Whitten\cite{jing92} who
made a two-determinant ansatz at the MCSCF level and perform
a CI calculation using this reference.\\
We want to present a multi-reference self-consistent field (MCSCF) 
calculation for the ideal and reconstructed Si(100) surface to cope
with the difficult bond structure. To deal with the dynamical correlations
we supplement it with a multi-reference configuration interaction (MRCI)
calculation. In addition we perform LDA and gradient corrected (GGA)
density functional calculations for a direct comparison.\\
The paper is organized as follows: In Sec.II we present the 
multi-reference methods and provide the technical details. 
The resulting wave function is discussed in Sec.III. In Sec.IV we present
the reconstruction energy, the geometry of the symmetric dimer
and the question of dimer buckling. Conclusion follows in Sec.V.

\section{Multi-reference methods}

Only closed-shell systems or high-spin open-shell system can be 
described within a single-determinant approach such as Hartree-Fock or
single-reference configuration interaction (CI) methods to 
include correlations.
In all other cases the suitable method is a
multi-configuration self-consistent field (MCSCF) approach.
The wave function is written as
\begin{equation}
|\Psi_{\rm MCSCF}> = \sum_{I}c_I |\Psi_I>,
\end{equation}
where $|\Psi_I>$ is a determinant built up from molecular orbitals and
$c_I$ are the so-called CI-coefficients. The difference to a 
single-reference CI approach is that not only the CI-coefficients are optimized
but also the coefficients of the orthonormal orbitals building up the $|\Psi_I>$. 
Therefore the number of determinants involved in the MCSCF should be
small compared with conventional CI. 
In the case of the Si(100) surface we select all possible determinants
combining the four singly occupied $sp^3$ hybrids to a singlet.
In addition we reoptimize the closed-shell orbitals of the first layer
to account for the influence of the reconstruction on the bulk.
The electronic correlations treated by a MCSCF ansatz are often called
static correlations.\\ 
To describe the dynamical correlations we apply on top of the MCSCF
a multi-reference configuration interaction (MRCI) with single and double 
excitations.
\begin{equation}
|\Psi_{\rm MRCI}> = \left( 1+\sum_{i\mu}\eta_{i}^{\mu}c^{\dagger}_i
c_{\mu} + \sum_{ij\mu\nu}\eta_{ij}^{\mu\nu}c^{\dagger}_i c^{\dagger}_j
c_{\mu}c_{\nu}\right)  |\Psi_{\rm MCSCF}>,
\end{equation}
where $\mu$ and $\nu$ numbers the occupied orbitals and 
$i$ and $j$ the unoccupied ones. 
The coefficients $\eta_{i}^{\mu}$ and $\eta_{ij}^{\mu\nu}$
are determined variationally in the MRCI procedure.
We restrict the dynamically correlated 
space to the surface orbitals.\\
We perform all calculations with the quantum-chemical program
package MOLPRO94\cite{molpro}, only for the DFT-calculations 
we use the latest version MOLPRO96\cite{molpro}.
\begin{figure}
\psfig{figure=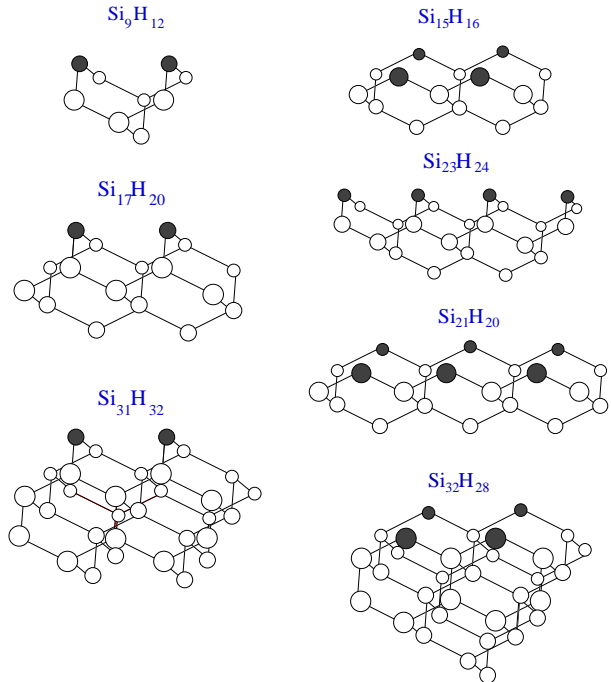,width=8cm}
\caption{\label{cluster}The figure shows 
all Si$_{x}$H$_{y}$ clusters which are used
in the calculations; the surface atoms are shaded;
the hydrogen atoms are not drawn.}  
\end{figure}
The smallest possible cluster with one dimer is ${\rm Si}_9$, the 
bulk-like bonds are saturated with 12 hydrogens. To test the influence of the 
bulk layers we perform calculations on ${\rm Si}_{17}{\rm H}_{20}$ and
${\rm Si}_{31}{\rm H}_{32}$, where the number of bulk layers increases.
The Si---Si distance in the bulk is taken from experiment (2.352\AA ),
the distance to the saturating hydrogen atoms is chosen equal to 
the one in ${\rm SiH}_{4}$ (1.480\AA ). In test calculations the 
geometry of the symmetric reconstruction is chosen such that 
all bonds have the bond length of the bulk. To test the interaction of 
the dimers we consider a ${\rm Si}_{15}{\rm H}_{16}$ cluster which
contains two parallel dimers and ${\rm Si}_{23}{\rm H}_{24}$
which has two dimers in line. Additionally we used a ${\rm Si}_{21}{\rm H}_{20}$
cluster with three parallel dimers.
The final results are obtained from ${\rm Si}_{32}{\rm H}_{28}$, which 
simulates six bulk layers and two dimers. All clusters are
shown in Fig. \ref{cluster}.\\
We apply a 4-valence-electron scalar-relativistic energy-consistent 
pseudopotential for Si\cite{bergner93} with the corresponding valence 
basis set\cite{bergner93}.
It is supplemented by various sets of polarization functions,
which are optimized by performing correlated calculations for the free atom.
For hydrogen we apply a minimal basis consisting of the fully contracted 1$s$ functions
of Dunning\cite{dunning89}.

\section{Many-body wave-function of the ideal and reconstructed surface}

For the ideal surface we perform a high-spin ($S=2$) 
single-determinant Hartree-Fock calculation to generate the starting orbitals
for the MCSCF calculation in which the
four singly occupied orbitals in the active space are combined 
to a singlet ground-state. In addition the closed-shell
orbitals of the first bulk layer are reoptimized. One way to discuss the
resulting orbitals is to calculate the natural orbitals which
diagonalize the density matrix (see Table \ref{orbital}). 
We find three quite close-lying orbitals two
of which have binding and one has antibinding character. Even the 
second antibinding one has a mean occupation of
half an electron. 
\begin{table}
\caption{\label{orbital}The mean occupation
number of the natural orbitals and their binding character is shown 
both for the ideal and reconstructed
surface.}
\begin{tabular}{c||c|c|}
&character&$\bar{n}$\\
\hline
ideal&1. bind.&1.52\\
&2. bind.&1.06\\
&3. antib.&0.94\\
&4. antib.&0.48\\
\hline
sym. reconstr.&1. bind.&1.98\\
&2. bind.&1.63\\
&3. antib.&0.37\\
&4. antib.&0.02\\
\end{tabular}
\end{table}
Regarding the coefficients of the MCSCF expansion there are 
ten configurations with a coefficient greater than 0.1. This shows clearly that
the ideal surface can not be described within a single-determinant approach.
The singlet closed-shell HF energy lies about 1 eV higher in energy than the
corresponding MCSCF value.\\
For the symmetric reconstructed surface we start the MCSCF calculation
with the orbitals of a triplet
HF determinant. But the active space of the MCSCF calculation
is exactly the same as in the ideal case. The natural orbitals look quite 
different. A low-lying binding orbital is doubly occupied forming
a single bond between the surface atoms. A second binding orbital 
is occupied with 1.6 electrons.
A double occupancy of this orbital would correspond to a double-bond between
the surface atoms. The significant population of the first antibinding
orbital with 0.4 electrons contradicts the closed-shell picture.
The MCSCF expansion is dominated by the configuration where the two 
binding orbitals are occupied, but the occupation of
the first antibinding orbital is not negligible.
In a single-determinant description the triplet state would be energetically
favoured by 0.9 eV leading to an unphysical ferromagnetic surface.
In the MCSCF calculation the triplet state is about 0.3 eV higher in energy
than the singlet ground state.\\
In both cases, i.e., in the ideal and in the symmetrically reconstructed surface    
we find a singlet ground state which can be only described within a 
multi-configuration approach. Through this we can provide for both cases
an adequate description. So we can compare the ground-state energies
of the two structures yielding the reconstruction energy.
   
\section{Results and Discussion}

\subsection{Basis set dependence}
In a first step we want to test the quality of the basis set used.
We calculate the reconstruction energy on the MCSCF and MRCI level
for the ${\rm Si}_9{\rm H}_{12}$
cluster. Although the question of the basis is a technical one,
it is possible to draw some physical conclusions.\\
The default basis is of  valence-double-$\zeta$ quality 
($4s4p)/[2s2p]$ supplemented by one $d$ polarization function with
exponent 0.4. In Table \ref{basis} the results are
listed. Calculating the reconstruction energy without any $d$-function
yields only 70\% of the value with the default basis. Applying a $d$-function only 
\begin{table}
\caption{\label{basis}Influence of the atomic basis set on the 
reconstruction energy at different correlation levels. The calculations
are performed in the ${\rm Si}_9{\rm H}_{12}$ cluster.}
\begin{tabular}{c||c|c|}
basis&$\Delta E_{\rm MCSCF}$[eV]&$\Delta E_{\rm MRCI}$[eV]\\
\hline
$[2s2p]$&0.90&1.08\\
red. basis&1.17&1.54\\
$[2s2p]1d$&1.26&1.56\\
$(4s4p)1d$&1.27&1.59\\
$[3s3p]2d$&1.29&1.60\\
$[3s3p]2d1f$&1.33&1.73\\
$(4s4p)3d2f1g$&1.34&1.73\\
\end{tabular}
\end{table}
for the surface and the first layer atoms
(reduced basis) we determine 93\% of the default MCSCF energy 
and 97\% of the 
default MRCI energy. The better value of the MRCI is clear, because in this
calculation only the surface orbitals are involved, whereas in the
MCSCF the first layer is reoptimized, too. This basis set will be
our choice for the larger clusters. The increase of the reconstruction
energy when supplementing a $d$ polarisation function indicates the important 
role of the electronic correlations on the formation of the bond in the 
dimer. Uncontracting  totally the
valence basis set yields only a small increase in the
reconstruction energy. This shows that the $[2s2p]$ basis is of
equal quality for both geometries. 
The same holds true if we split the $d$-exponent into
two (0.23, 0.8). To receive a balanced basis we uncontract
the valence basis to $[3s3p]$.
Applying an additional $f$-function (exponent 0.35) the reconstruction energy 
rises by 6\% on the MCSCF level and by 10\% on the MRCI level. This indicates
that to correlate the $\pi$-bond
the $f$-function is more important in the reconstructed geometry, 
especially for the dynamical correlation covered by the MRCI.  
Uncontracting the valence basis set totally to $(4s4p)$ and adding a
$(3d2f1g)$ polarization set (exponents from Ref.[\onlinecite{steinbrenner94}])
yields nearly no changes (smaller than 1\%) on the reconstruction
energy on the MCSCF and MRCI level. This indicates that the basis set limit 
is reached for the reconstruction energy . 

\subsection{Influence of bulk layers}

To check the influence of the bulk layers on the surface reconstruction
we perform MCSCF and MRCI calculations for three different clusters using
the reduced basis. ${\rm Si}_{9}{\rm H}_{12}$ simulates
three bulk layers, ${\rm Si}_{17}{\rm H}_{20}$ four layers and 
${\rm Si}_{31}{\rm H}_{32}$ five layers whereas all have one surface dimer;
for results see Table \ref{layer}.
\begin{table}
\caption{\label{layer}Influence of the bulk layers on the reconstruction
energy at different correlation levels. The calculations
are performed in the reduced basis.}
\begin{tabular}{c||c|c|}
cluster&$\Delta E_{\rm MCSCF}$[eV]&$\Delta E_{\rm MRCI}$[eV]\\
\hline
${\rm Si}_9{\rm H}_{12}$&1.17&1.54\\
${\rm Si}_{17}{\rm H}_{20}$&1.33&1.67\\
${\rm Si}_{31}{\rm H}_{32}$&1.30&1.61\\
\end{tabular}
\end{table}
The changes between three and four layers are quite significant, i.e. up to 14\%
of the reconstruction energy at the MCSCF level. 
This shows that the ${\rm Si}_{9}{\rm H}_{12}$
cluster is not fully reliable for describing the surface. 
The difference between
four and five layers is up to 4\% and is tolerable for the 
determination of the reconstruction energy. But in our opinion it is not possible to 
calculate reliable relaxed distances of the bulk atoms, which is
a shortcoming of the cluster approach. Experimentally it is 
proven\cite{felici97} that
there are relaxations up to the fifth bulk layer.   

\subsection{Interaction of dimers}

Although we have checked in the last section, that the 3-layer clusters
can not fully describe the surface reconstruction we select
four different 3-layer clusters in order to test the interaction of the dimers.
${\rm Si}_{9}{\rm H}_{12}$ has one dimer, ${\rm Si}_{15}{\rm H}_{16}$
has two parallel dimers and ${\rm Si}_{21}{\rm H}_{20}$ has three.
${\rm Si}_{23}{\rm H}_{24}$ has two dimers in line.
\begin{table}
\caption{\label{dimer}Influence of the dimer interaction 
on the reconstruction
energy at different correlation levels. The calculations are 
performed in the reduced basis.}
\begin{tabular}{c||c|c|c|c}
cluster&$\Delta E_{\rm MCSCF}$[eV]&$\Delta E_{\rm MRCI}$[eV]&
$\Delta E_{\rm LDA}$[eV]&$\Delta E_{\rm GGA}$[eV]\\
\hline
${\rm Si}_9{\rm H}_{12}$&1.17&1.54&2.01&1.43\\
${\rm Si}_{15}{\rm H}_{16}$&1.15&1.46&1.85&1.27\\
${\rm Si}_{21}{\rm H}_{20}$&1.13&---&1.76&1.27\\
\hline
${\rm Si}_{23}{\rm H}_{24}$&1.15&1.46&---&---\\
\end{tabular}
\end{table}
Concerning the reconstruction energy on the MCSCF level, all
selected clusters differ by at most 4\% concluding that
the interaction on the MCSCF level is weak (see Table \ref{dimer}). 
On the MRCI level
we calculate only the cluster with one and two parallel dimers.
The difference is larger than at the MCSCF level indicating that 
the dynamical correlations are
more important for the dimer interaction than are the static correlations.\\
Treating the clusters in a LDA approach (Slater-Dirac exchange\cite{slater74} and
Vosko-Wilk-Nusair correlation functional\cite{vosko80}) the interaction between 
the dimers is larger by up to 12\%. This can be understood 
from the wave function structure resulting from the LDA. 
If more than one dimer exist, there is an energy gain due to 
delocalization of the surface electrons. 
In a gradient corrected density functional approach (GGA, Becke exchange
\cite{becke88} and Lee-Yang-Parr correlation functional\cite{lee88})
we found a similar behaviour, but there the effect of delocalization is 
smaller than in the LDA and occurs only between two dimers.  
This delocalization energy is larger 
on the ideal surface than on the reconstructed surface due to the
close lying and therefore well mixing states. This results in
an overestimation of the dimer interaction on the LDA level. If the surface
is treated on the MCSCF level the mixing between different states 
can occur within one dimer, so that both ideal and reconstructed 
surface are treated at the same level of accuracy.    

\subsection{Surface reconstruction energy --- comparison of different methods}

Combining the results of the previous sections we select
the ${\rm Si}_{32}{\rm H}_{28}$ cluster for our final calculation.
This cluster has two dimers and the bulk is modelled up to the 
sixth bulk layer. Due to the quite large size of the cluster
we can treat it only with the reduced basis. The surface reconstruction
energy at the MCSCF level is 1.13 eV. Including dynamical correlations
yields an increase of nearly 40\% resulting in 1.56 eV. Concerning the limited
basis used in this calculation we can estimate based on our basis
set test calculations a final value of 1.75 eV. The error due to the 
cluster size is estimated to be by $\pm$5\% . Comparing our result with 
various LDA-slab calculations 
(Dabrowski et al.\cite{dabrowski92}
1.5 eV; Fritsch et al.\cite{fritsch95} 1.51 eV;  
Ramstad et al.\cite{ramstad95} 1.8 eV; Roberts et al.\cite{roberts90}
2.02 eV)
the agreement is remarkably good.   
In order to compare the results of different methods in more detail we 
apply the different methods to the same cluster
with the same basis set. We choose again the ${\rm Si}_{9}{\rm H}_{12}$ cluster 
using the $[2s2p]1d$ basis. DFT methods are applied with two different
exchange correlation functions. The LDA described above yield a
reconstruction energy of 2.01 eV which is nearly 30\% above the MRCI value
of 1.56 eV. This value is certainly lowered due to dimer 
interaction as described in the previous section. 
Using the GGA approach 
the reconstruction energy (1.43 eV) is underestimated by 8\% .
The DFT treatments yield a singlet ground state for the reconstructed surface 
in contradiction to a
single-determinant Hartree-Fock calculation,
but the reconstruction energy is dependent on the
chosen functional. The results show the known feature
that LDA overestimates binding whereas the GGA gives a substantial 
improvement.
 
\subsection{Geometry of the symmetric reconstruction}

We argued in the previous sections that the clusters are too small
for treating the relaxations of the bulk layers due to surface 
reconstruction. But it is reasonable to optimize the dimer bond
length ($b_{\rm dimer}$) and the bond length of the
surface atoms to the first layer atoms ($b_{\rm 1.layer}$);
for results see Table \ref{lattsym}. To check the
cluster-size dependence we perform the calculations for the 
${\rm Si}_{9}{\rm H}_{12}$ cluster as well as for the 
${\rm Si}_{17}{\rm H}_{20}$ cluster within the $[2s2p]1d$ basis.
The results show, that the bond lengths are only slightly 
cluster dependent. The changes are less than 0.02\AA .
\begin{table}
\caption{\label{lattsym}Optimized dimer bond length and 
first layer bond length of the symmetric reconstructed 
surface.} 
\begin{tabular}{cc||c|c}
method&&$b_{\rm dimer}$[\AA ]&$b_{\rm 1.layer}$[\AA ]\\
\hline
MCSCF&${\rm Si}_9{\rm H}_{12}$&2.41&2.36\\
MRCI&${\rm Si}_9{\rm H}_{12}$&2.37&2.36\\
\hline
MCSCF&${\rm Si}_{17}{\rm H}_{20}$&2.40&2.36\\
MRCI&${\rm Si}_{17}{\rm H}_{20}$&2.35&2.36\\
\hline
LDA&[\onlinecite{roberts90}]&2.23&2.29\\
&[\onlinecite{ramstad95}]&2.23&2.27\\
&[\onlinecite{fritsch95}]&2.26&2.31\\
\end{tabular}
\end{table}
The dynamical correlations reduce the dimer bond length by about
2\% . Compared with
the dimer bond-length of the asymmetric dimer from different experiments
(Bullock et al.\cite{bullock95} 2.25\AA ; 
Jedrecy et al.\cite{jedrecy90} 2.32\AA ; Aono et al.\cite{aono82}
2.4$\pm$0.1\AA ; Holland et al.\cite{holland84} 2.47\AA ;
Yang et al.\cite{yang83} 2.54\AA ; Felici et al.\cite{felici97}
2.67\AA ) we agree quite well with the second and third measurement.
LDA-slab calculations (see Table \ref{lattsym}) 
yield a dimer bond length quite below the bulk value.
   
\subsection{Asymmetric reconstruction}

\begin{figure}
\psfig{figure=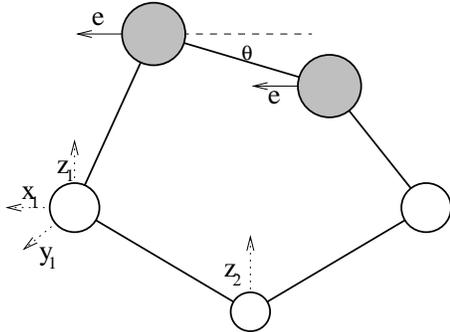,width=6cm,angle=-90}
\caption{\label{asym}The figure shows
a projection of the Si$_{9}$H$_{12}$ cluster where
the displacements concerned for the asymmetric buckling
are indicated.}
\end{figure}

In this section we want to arise the question of the dimer buckling.
Up to now we allowed only for the symmetric reconstruction which
maintains the local C$_{\rm 2v}$-symmetry. We test the buckling
by lowering the symmetry in the ${\rm Si}_{9}{\rm H}_{12}$ cluster
with the $[2s2p]1d$ basis. We allow for a buckling angle $\Theta$
and a displacement of the dimer parallel to the surface (see Fig. \ref{asym}). 
The dimer
bond length is fixed to the value obtained in the optimization
of the symmetric case. In the MCSCF and MRCI treatment
the symmetric dimer has the lowest energy contrary to
the DFT calculations where we 
find a buckling of 9$^o$ in LDA and of 8$^o$ in GGA.
The energy gain is very small, i.e., 0.031 eV in LDA and
0.014 eV in GGA. From this we can conclude that the 
used approximations to DFT
overestimate the electronic contribution to the
buckling  due to an overestimation of the closed-shell
structure.\\
The experimentally observed buckling could 
be driven by relaxation of the bulk atoms.
Although the ${\rm Si}_{9}{\rm H}_{12}$ cluster is quite small to resolve 
relaxations we performed some test calculations within this model.
We allow for the first bulk layer atoms to relax in all directions,
whereas we regard for the second bulk layer only a relaxation
perpendicular to the surface (see Fig. \ref{asym}). On the MCSCF level, where        
all orbitals are reoptimized we do not find any buckling although 
there is a significant relaxation of the first bulk layer atoms in
the dimer direction of about 0.08\AA\ and of the second bulk layer 
atoms perpendicular to the surface of about 0.09\AA . 
The other relaxations are smaller by
about one order of magnitude. From this we can conclude that in our cluster 
model the bulk relaxations do not drive the asymmetric buckling.\\
To account for the dimer interaction we regard the ${\rm Si}_{15}{\rm H}_{16}$ cluster
with two dimers. We model the asymmetric p(1$\times$2) structure as well as
the  p(2$\times$2) structure with alternating buckled dimers. For fixed buckling
angle the p(2$\times$2) structure is lower in energy than the asymmetric p(1$\times$2) structure,
e.g., for an angle of 2$^o$ by 0.018 eV per dimer and for an angle
of 10$^o$ by 0.060 eV. But still the symmetric dimer is lowest in energy, also if we allow
for the relaxations we determined in the ${\rm Si}_{9}{\rm H}_{12}$ cluster. This shows that
the interaction of two dimers is not enough to stabilize a buckled structure.
This is in contradiction to the LDA cluster results from Yang et al.\cite{yang97}
who find a stable  p(2$\times$2) structure  in the ${\rm Si}_{15}{\rm H}_{16}$ cluster.\\
Our finding of a symmetric ground state is in agreement with all previous correlation
calculations, which show that the proper treatment of the electronic correlations
stabilize the symmetric reconstruction. We can exclude an asymmetric  p(1$\times$2) structure,
which LDA slab calculations\cite{roberts90,dabrowski92,fritsch95,ramstad95}
find about 0.1eV more stable than the symmetric 
one. As we can only calculate the interaction of two dimers, there is the possibility
that the interaction of many dimers can yield a buckled ground state.
But regarding the STM experiment of Badt et al.\cite{badt94}, who find a 
buckled c(2$\times$4) structure at 90K, they claim that the buckling can 
be induced by a defect or a step at the surface. The defect-free surface
is difficult to prepare, so the question of buckling of the Si(100) surface is 
experimentally still not solved.\\
To address the question of metallicity of the symmetric reconstructed 
Si(100) surface, which is found in various DFT calculations and discussed
in detail in Ref.[\onlinecite{rohlfing95}], in a many-body description
as applied here there can exist a gap due to electronic correlations
in the symmetric reconstruction, too. But in principle 
the question of metallicity cannot be solved in a cluster approach.

\section{Conclusion}

We have performed MCSCF calculations for the ideal and
reconstructed Si(100) surface. In both cases the 
many-body ground state is a singlet, but it can only be
described by several determinants. 
On the reconstructed surface a dimer is formed. The
bond structure is a mixture between a double-bond and a 
single-bond with two singly occupied orbitals. 
To account for the dynamical 
correlations we apply on top of the MCSCF result a MRCI
calculation. Determining the reconstruction energy
we find good agreement with various LDA slab calculations.
The dimer bond length agrees better with experiment than the LDA one.
The asymmetric reconstruction is overestimated in the LDA
while our fully correlated treatment finds the symmetric dimer more
stable regarding the electronic structure of one dimer and relaxations of 
the bulk atoms.
The buckling observed in experiment can be due to a small concentration of
defects on the Si(100) surface. It is possible that an alternating buckled 
structure is induced by defects and persist
down to low temperatures due to dimer interactions. 

\section{Acknowledgements}

I gratefully acknowlegde the valuable scientific discussions with
and suggestions of Prof.~H.~Stoll, Stuttgart, and Prof.~P.~Fulde,
Dresden. I also thank Dr M.~Dolg and F.~Schautz for the
support concerning the multi-reference calculations.

\end{document}